\documentclass[prd,showpacs,preprintnumbers,twocolumn,superscriptaddress,floatfix]{revtex4}
\usepackage[utf8]{inputenc}
\usepackage{bm,amsmath,amssymb}
\usepackage{graphicx}
\usepackage{subfigure}
\usepackage{color}
\usepackage{hyperref}
\usepackage{multirow}
\usepackage{soul}
\usepackage{comment}
\usepackage{url}

\let\oldalign\align
\def\align{\linenomath\oldalign}

\begin{document}

\title{Eigenvalues of the QCD Dirac matrix with improved staggered quarks in the continuum limit  }

\author{Olaf Kaczmarek}
\affiliation{Fakult\"at f\"ur Physik, Universit\"at Bielefeld, D-33615 Bielefeld,
Germany}

\author{Ravi Shanker}
\email{rshanker@imsc.res.in}
\affiliation{The Institute of Mathematical Sciences, a CI of 
Homi Bhabha National Institute, Chennai,  600113, India}

\author{Sayantan Sharma}
\affiliation{The Institute of Mathematical Sciences, a CI of 
Homi Bhabha National Institute, Chennai,  600113, India}

\begin{abstract}

We calculate the eigenmodes of the Highly Improved Staggered Quark 
(HISQ) matrix near the chiral crossover transition in QCD with 
$2+1$ flavors with the aim to gain more insights into its temperature 
dependence. On performing the continuum extrapolation, we do not 
observe any gap opening up in the infrared part of the eigenvalue density 
of the QCD Dirac operator; instead we observe a peak. The existence of the 
peak and oscillations of the infrared eigenmodes can be understood 
in terms of an interacting ensemble of instantons. From the properties of 
the continuum extrapolated eigenspectrum we further 
show that the anomalous $U_A(1)$ part of the chiral symmetry is not 
effectively restored simultaneously along with its non-singlet counterpart. 
We provide an explanation for this observation, further showing interesting 
connections between the anomalous $U_A(1)$ restoration and the change in the
infrared part of the eigenvalue distribution.

\end{abstract}

\maketitle


\section{Introduction}
The eigenvalue spectrum of the quark Dirac operator contains valuable information 
about the fundamental properties of Quantum Chromodynamics (QCD). The chiral condensate 
which acts as an (pseudo) order parameter for the chiral (crossover) transition in 
QCD is related to the density of near-zero eigenvalues~\cite{Banks:1979yr}. In fact 
it was shown from very general considerations that the formation of the chiral 
condensate is related to the occurrence of small eigenvalues that scale proportional 
to the volume~\cite{Leutwyler:1992yt}. The breaking of the non-singlet 
part of chiral symmetry, i.e., $SU_A(2)\times SU_V(2)\to SU_V(2) $ of QCD with 
physical quark masses at the crossover temperature $T_c=156.5\pm 1.5$
MeV~\cite{HotQCD:2018pds} can also be explained in terms of modifications
in the deep infrared part of the eigenvalue density. 
The flavor-singlet $U_A(1)$ part of the chiral symmetry 
on the other hand, is anomalous yet is believed to play an 
important role in determining the nature of the chiral phase
transition~\cite{Pisarski:1983ms,Butti:2003nu,Pelissetto:2013hqa}. The temperature dependence 
of the amount of $U_A(1)$ breaking near the chiral crossover 
transition in QCD can be only determined using non-perturbative 
lattice techniques and is a topic of contemporary interest 
in lattice QCD; see, for e.g., Ref.~\cite{Sharma:2018y2,Lombardo:2020bvn} 
for recent reviews. Whereas there are some very compelling evidence 
that show $U_A(1)$ remains effectively broken in $2+1$ flavor QCD with 
physical quark mass $m$
~\cite{HotQCD:2012vvd,Buchoff:2013nra,Bhattacharya:2014ara,Dick:2015twa,Petreczky:2016vrs,Bazavov:2019www,Ding:2020xlj}, 
even when $m\to 0$~\cite{Kaczmarek:2021ser}, there are lattice studies which also favor 
an effective restoration at $T_c$~
\cite{Cossu:2013uua,Chiu:2014Ld,Tomiya:2016jwr,Brandt:2016daq,Aoki:2020noz,Aoki:2021qws}.

The eigenvalue spectrum of the QCD Dirac matrix also encodes within it some remarkable universal 
properties. It was shown that the route toward achieving the thermodynamic limit for the infrared
modes of the Dirac operator is universal~\cite{Shuryak:1992pi}, for any number of light
quark flavors. The existence of a non-zero chiral condensate leads to a sum rule involving the sum 
of inverse squares of these small eigenvalues~\cite{Leutwyler:1992yt}. These sum rules are universal 
irrespective of the details of the nature and type of gauge interactions~\cite{Shuryak:1992pi,Verbaarschot:1993pm}
and could be derived from chiral random matrix theory~\cite{Verbaarschot:1994qf}. A good  
agreement was demonstrated for the distribution of the small eigenvalues and the spectral 
density of lattice QCD Dirac operator and chiral random matrix theory at zero temperature 
on small lattice volumes~\cite{Berbenni-Bitsch:1997zmi}. In fact universal correlations between 
higher order spectral functions in a random matrix theory has been derived~\cite{Akemann:1996vr}, 
and its connection to QCD was discussed. At finite temperature the universal features of infrared 
eigenvalues can be also accounted for within a random matrix 
theory~\cite{Jackson:1997ud,Garcia-Garcia:2000lra,Akemann:2021yim}.
Additionally the infrared eigenvalue spectrum of QCD has more subtle features. A near-zero peak of 
localized eigenvalues has been observed for finite lattices, mixing with but very different from 
the delocalized bulk modes whose spectral density follows random matrix 
statistics~\cite{Sharma:2018y2,Giordano:2021qav}. Whether or not such a feature survives in the 
continuum limit is yet to be ascertained. Previous studies of quark Dirac spectrum in an instanton 
liquid ensemble~\cite{Verbaarschot:1994te,Garcia-Garcia:2000lra} at zero temperature have observed 
a similar peak-like feature.

With increasing temperature the localized modes start separating out from the random 
bulk modes leading to the opening up of a mobility edge~\cite{Giordano:2021qav}. The 
corresponding temperature where a finite mobility edge separates the bulk modes from the 
localized one was initially estimated from lattice studies to be identical to $T_c$ in dynamical~
\cite{Garcia-Garcia:2006vlk,Kovacs:2010wx,Kovacs:2012zq,Giordano:2013taa,Ujfalusi:2015nha,Giordano:2016nuu,Holicki:2018sms,Cardinali:2021fpu,Kehr:2023wrs} 
as well as in quenched QCD~\cite{Kovacs:2017uiz,Vig:2020pgq}, reminiscent of an  Anderson-like transition that 
is observed in disordered semi-metals~\cite{Anderson:1958vr}. However independent lattice studies do 
discuss another possible scenario where the opening of a finite mobility edge may occur at temperatures 
higher than $T_c$~\cite{Alexandru:2021xoi}, with an intermediate phase consisting of scale-invariant 
infinitely extended infrared modes~\cite{Alexandru:2019gdm,Alexandru:2021pap} strongly interacting with 
the bulk modes leading to a singularity at the mobility edge. 

Most of the previous lattice QCD studies were either performed in the quenched limit 
or with dynamical quarks but away from the physical point and for finite lattice spacings. 
On a finite lattice, the most often used lattice discretization, i.e., the staggered fermions 
only has a remnant of the continuum chiral symmetry group due to mixing of spin and flavor 
degrees of freedom. Furthermore the anomalous part of the chiral symmetry in the continuum 
is not realized exactly by the staggered/Wilson quarks and is expected to be recovered only in 
the continuum limit. We, for the first time study the properties of the eigenvalue spectrum 
of (highly) improved dynamical staggered Dirac operator in large volume lattices by carefully performing 
a continuum extrapolation. We show that the deep infrared spectrum of the QCD Dirac operator has indeed a 
peak of near-zero modes which survives in continuum. These are distinct from other infrared modes which 
have a linearly rising density and a quadratic level repulsion similar to a certain class of random matrix 
theories. These so-called bulk modes are delocalized in volume as compared with the near-zero modes, 
and they tend to distinctly disentangle from each other at a temperature $\sim 1.15~ T_c$, which is 
also where $U_A(1)$ is \emph{effectively} restored. In the subsequent sections we discuss our results 
and also provide a unified physical explanation of these phenomena we observe.


\section{Numerical Details}
In this work we use the gauge configurations for $2+1$ flavor QCD with physical 
quark masses generated by the HotQCD collaboration using Highly Improved Staggered 
quark (HISQ) discretization for the fermions and 
tree-level Symanzik improved
gauge action. These ensembles 
have been previously used to measure the equation of state of QCD both at zero and 
finite baryon density~\cite{Bazavov:2017dus,HotQCD:2018pds}. The Goldstone pion mass is
set to $140$ MeV, and the kaon mass is $435$ MeV for these configurations.  We 
focus on five different temperatures, one below $T_c$ and others above $T_c$. For 
most of these temperatures we consider three different lattice spacings corresponding 
to $N_\tau=8, 12, 16$, the details of which are mentioned in Table~\ref{tab:table1}.
The number of spatial lattice sites was chosen to be $N_s=4N_\tau$ such 
that the spatial volume in each case was about $4$ fm, which ensures that the system 
is close to the thermodynamic limit. We calculated the first $60, 100, 200$ eigenvalues 
of the massless HISQ Dirac matrix for $N_\tau=16, 12, 8$ respectively on these gauge ensembles 
using conjugate gradient method based algorithms. We have fixed the bin size $\lambda a = 0.001$ 
for each $N_\tau$  for measuring the eigenvalue density and performed a jack-knife analysis to 
remove any auto-correlation effects among the data in the bins.

\begin{table}[ht]
 \centering
 \begin{tabular}{|c|r|r|r|r|}
 \hline \hline
 $T$ (MeV) & $\beta$ & $N_s$ & $N_\tau$ & $N_{confs}$  \\
 \hline
  145 & 6.285 & 48 & 12 &1530\\
   145 & 7.010 & 64 &16  & 2860\\
  \hline
 \hline
 162 & 6.423 & 32 & 8 & 250 \\
  162 & 6.825 & 48 & 12 &1960\\
   162 & 7.130 & 64 &16  & 3390\\
  \hline
  166 & 6.445 & 32  & 8  & 400\\
 166 & 6.850 & 48  & 12  &2100\\
 166 & 7.156 & 64 & 16   &2190\\
 \hline
171 & 6.474 & 32 & 8  & 280\\
171 & 6.880 & 48 & 12  &1980\\
171 & 7.188 & 64 &16 &1040\\
 \hline
  \hline
176 & 6.500 & 32 & 8  & 240\\
176 & 6.910 & 48 & 12  & 330\\
\hline
  \end{tabular}
  \caption{Number of configurations for different temperatures ($\beta$ values) and the corresponding lattice sizes used in this work.}
  \label{tab:table1}
\end{table}

\section{Results}

\subsection{General features of the eigenvalue spectrum of QCD using HISQ Dirac operator 
in continuum limit}
In this section we study in detail the eigenvalue density $\rho(\lambda)$ of the 
quark Dirac operator in $2+1$ flavor QCD by performing a continuum extrapolation of the parameters 
characterizing the eigenspectrum calculated on the lattice with HISQ discretization.

\begin{figure*}[t]
\includegraphics[scale=0.46]{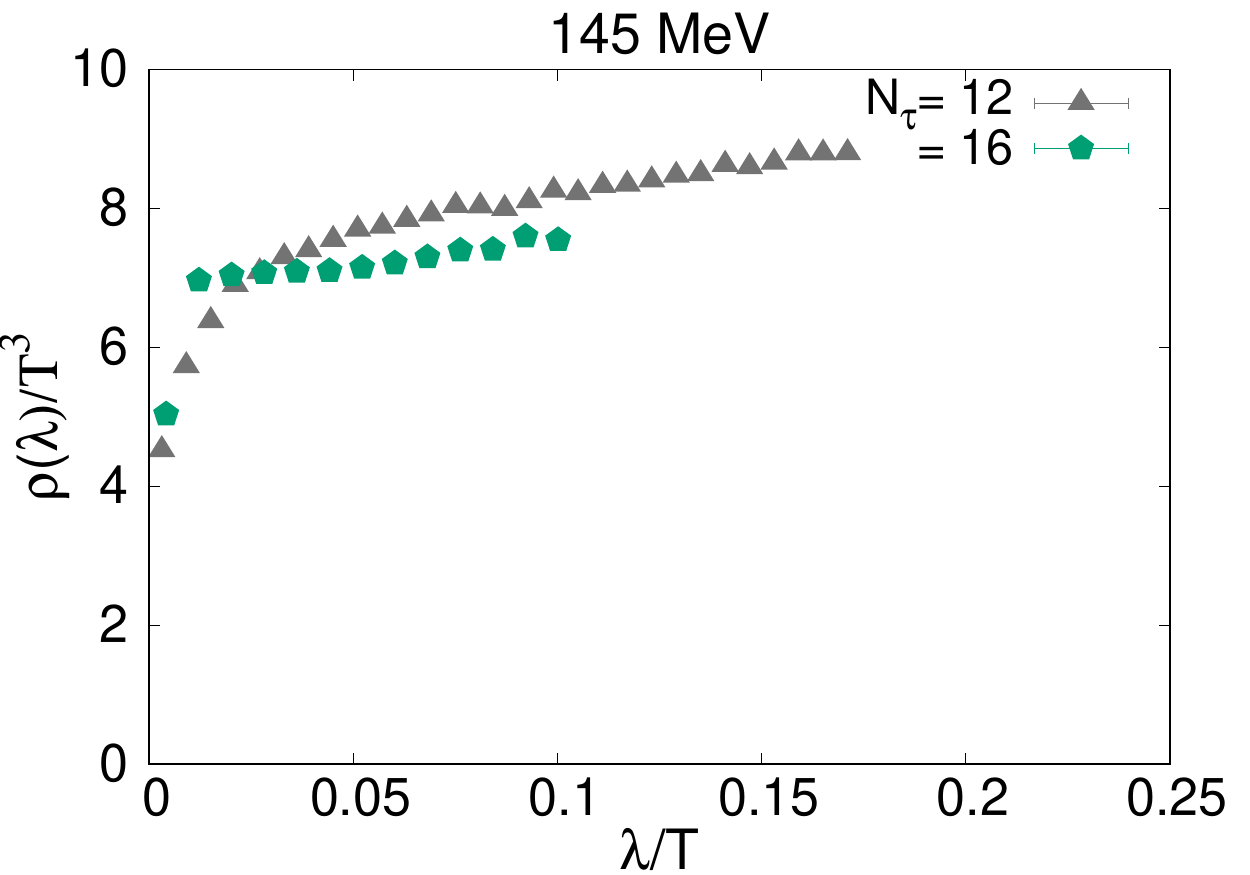}
\includegraphics[scale=0.46]{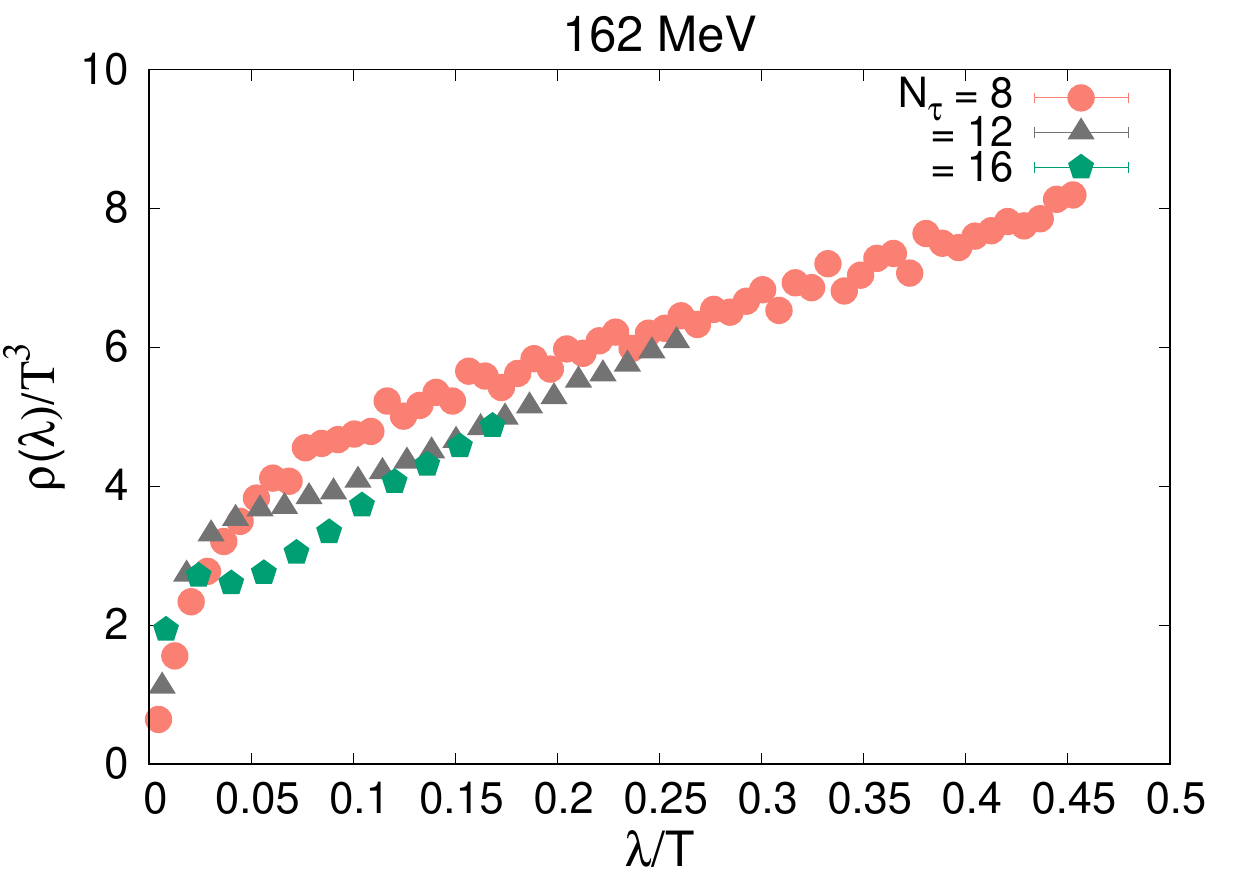}
\includegraphics[scale=0.46]{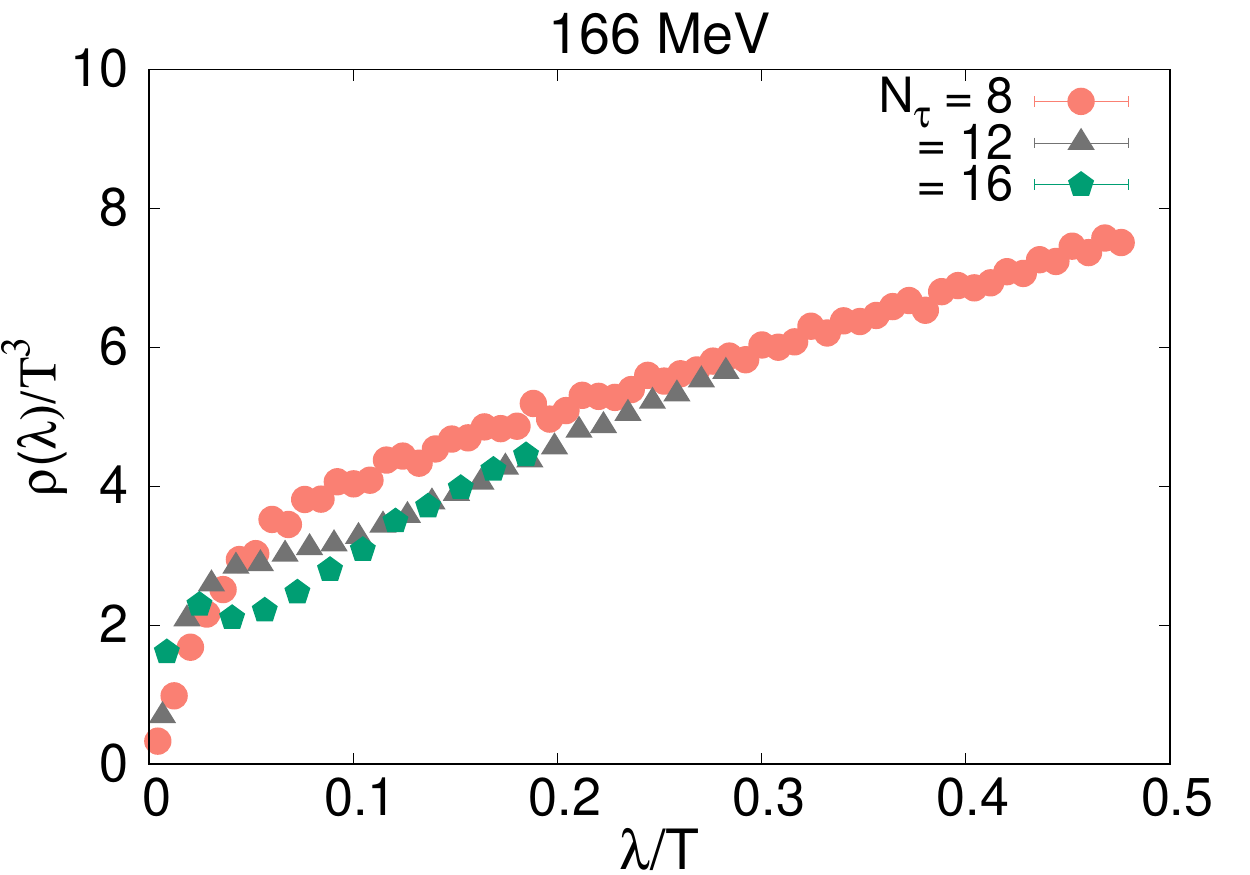}
\includegraphics[scale=0.46]{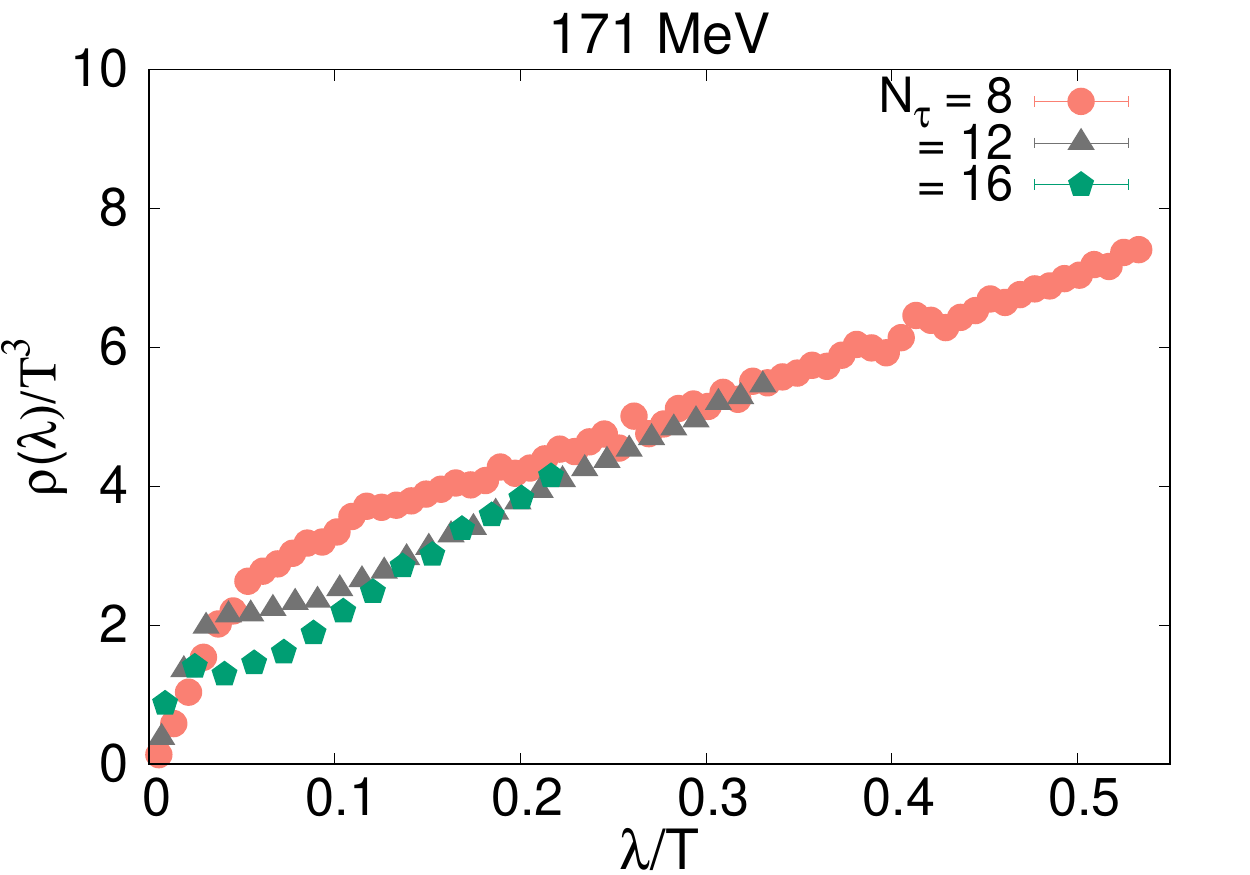}
\includegraphics[scale=0.46]{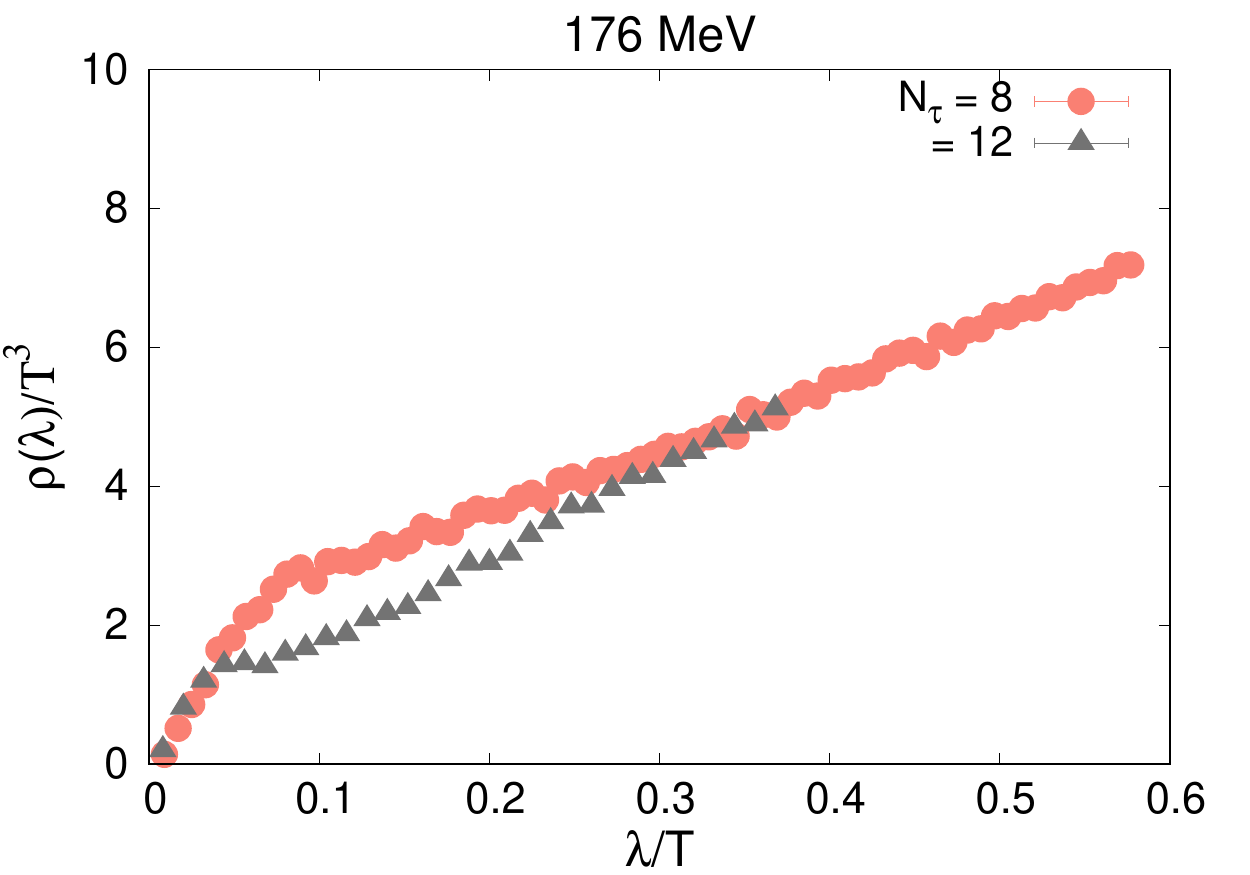}
\caption{Eigenvalue spectrum for HISQ Dirac operator for three different lattice spacings 
corresponding to $N_\tau=8, 12, 16 $ at $T= 162, 166, 171$ MeV , for two different 
lattice spacings, $N_\tau = 12, 16$ at $T=145$ MeV and $N_\tau = 8, 12$ at $T=176$ MeV. }
\label{fig:hisqev}
\end{figure*}
At zero temperature it is known from chiral perturbation theory~\cite{Smilga:1993in} that 
the linearly rising part of the eigenvalue density, due to the so-called bulk modes, is expressed as
\begin{equation}
\label{eqn:evzT}
\rho(\lambda) = \frac{\langle0\vert\bar \psi \psi\vert 0\rangle}{\pi}+
\vert \lambda\vert \langle0\vert\bar \psi \psi\vert 0\rangle^2 \frac{N_f^2-4}{32\pi^2N_fF_\pi^4}+..
\end{equation}
The intercept of bulk eigenvalue density gives the chiral 
condensate. The ratio of the slope and the intercept of the density as a function 
of $\lambda$ should be proportional to the chiral condensate.  We first focus on the 
intercept and the slope (linear in $\lambda$) of the eigenvalue density at the lowest 
temperature $T=145$ MeV, shown in the top left panel of Fig.~\ref{fig:hisqev}, and compare 
with the expectations from Eq.~\ref{eqn:evzT}. At this temperature we could only obtain a 
continuum estimate of the slope and intercept as we have data for two lattice spacings. 
From the continuum estimate of the intercept we obtain a chiral condensate 
$\langle0\vert\bar \psi \psi\vert 0\rangle/T^3=18.4$ using Eq.~\ref{eqn:evzT}. 
From the slope we could similarly extract the square $\langle0\vert\bar \psi \psi\vert 0\rangle^2$ 
and by substituting $N_f=3, F_\pi=94.14$ MeV the chiral condensate (normalized by 
$T^3$) to be $17.3$ which is consistent with the one extracted from the intercept.
 This demonstrates the consistency of our fit procedure. The value 
obtained here from the eigenvalue spectrum is also consistent with the value of 
$\langle0\vert\bar \psi \psi\vert 0\rangle/T^3=18.8$ obtained from the inversion 
of the HISQ Dirac operator on stochastic noise vectors and performing a continuum 
estimate using the $N_\tau=12, 16$ data on a much larger set of HotQCD 
configurations~\cite{Steinbrecher:2018jbv}. 
Thus we conclude here that the leading features of the eigenvalue density of QCD at 
$145$ MeV are indeed very well represented within chiral perturbation theory.


\begin{figure}[h]
\begin{center}
\includegraphics[scale=0.6]{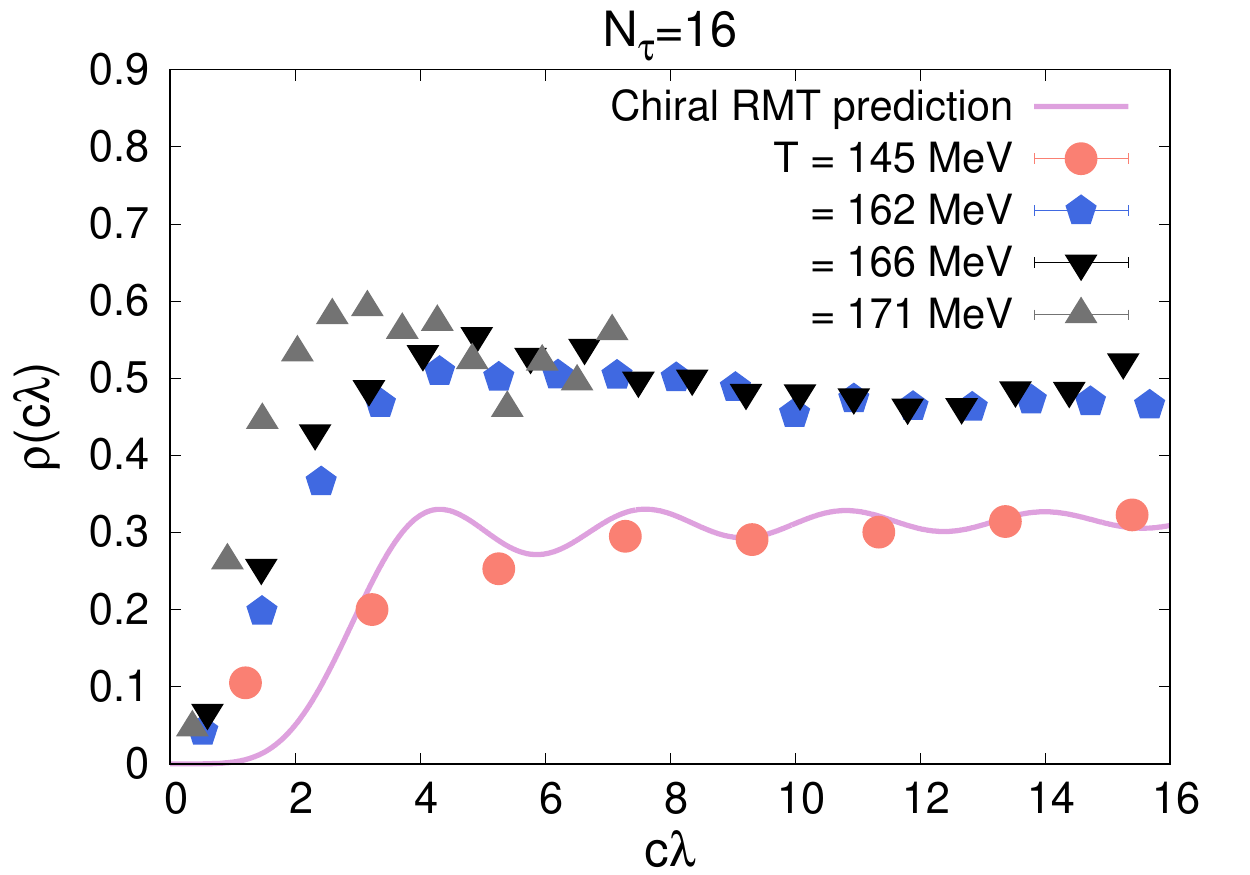}
\caption{Near-zero (scaled) eigenvalue density for HISQ Dirac operator at $T = 145, 162, 166, 171$ MeV for 
the finest lattice spacing corresponding to $N_\tau =16$ and its comparison with cRMT prediction 
available at $T=0$.  }
\label{fig:ILM}
\end{center}
\end{figure}

The bulk eigenvalue density in the chirally symmetric phase has been studied earlier in 
Ref.~\cite{Aoki:2012yj}. Most generally, it can be expressed as a function of $\lambda$ as
\begin{equation}
\label{eqn:evfit}
\frac{\rho(\lambda)}{T^3} =\frac{\rho_0}{T^3}+ \frac{\lambda}{T}.\frac{c_1(T,m)}{T^2}+~
\frac{\lambda^2}{T^2}.\frac{c_2(T,m)}{T}+\frac{\lambda^3}{T^3}c_3(T,m)~.
\end{equation}
Here $c_1$ is the coefficient that characterizes the leading-order growth of the 
eigenvalue spectrum in the deep infrared and $c_2$ is its next-to leading 
order coefficient which eventually has a $\lambda^3$ dependence predicted 
from perturbation theory. The intercept  $\rho_0$ gives the the chiral condensate. 
The coefficients $c_{1,2,3}$ can in general be a function of the temperature $T$ 
and the light-quark mass $m$. 

The results of the eigenvalue density $\rho(\lambda)/T^3$ as a function of $\lambda$ for $T>T_c$
are shown in Fig.~\ref{fig:hisqev}. On the finest available $N_\tau=16$ 
lattice, we observe two distinct features in the eigenvalue spectrum, a peak of near-zero eigenvalues 
and the linearly rising part, due to the bulk modes, as previously mentioned.  For $T\lesssim T_c$, 
the near-zero and the bulk eigenvalues overlap strongly making it impossible to distinguish them apart. 
At higher temperatures, the bulk eigenvalues separate out from the deep-infrared part of the spectrum 
allowing for near-zero modes to be distinctly visible. Comparing the results of different lattice 
spacings, we observe the same trend at each temperature above $T_c$, i.e., near-zero peak gets smeared 
with the bulk for coarser lattices and becomes more prominent in the continuum limit. This is thus 
a physical feature of the eigenspectrum and not a lattice artifact. In order to interpret its 
origin we recall that in the chiral random matrix theory (cRMT) at zero temperature, the scaled 
eigenvalue $(c\lambda)$ density of the Dirac operator for $N_f=2$ flavors and zero topological 
charge sector is distributed according to~\cite{Akemann:2016keq},

\begin{equation}
\label{eqn:liq}
\rho(c\lambda) = \frac{c\lambda}{2}\left[ J^2_2(c\lambda)-J_3(c\lambda)J_1(c\lambda) \right]~.
\end{equation}
To compare our data with the above formula, we take $c=V\langle0\vert\bar \psi \psi\vert 0\rangle/T$,
where $V$ is the spatial volume of the system, and the value of $\langle \bar{\psi} \psi \rangle$ at 
finite temperature is obtained from Ref.~\cite{Steinbrecher:2018jbv} which uses the same HotQCD gauge
configurations, a subset of which is used in this work. Further we also scale the eigenvalues such that
the first moment of probability distribution for lowest eigenvalues for the data matches with 
the first moment of Eq.~\ref{eqn:rmtzmode}. A comparison of near zero modes for four different 
temperatures, $T= 145, 162, 166, 171$ MeV, is shown in Fig.~\ref{fig:ILM}. We observe a good 
agreement with cRMT for $T= 145$ MeV, in particular, the initial few oscillations of the small 
eigenvalue density as a function of $c \lambda$. Incidentally an agreement between Eq.~\ref{eqn:liq}
and the eigenvalue density from the instanton liquid model (ILM) at $T=0$ was observed in~\cite{Verbaarschot:1996qm}.
However at finite temperature the oscillations in the eigenvalue density within ILM are smeared out over
a length scale $\sim 1/T $  ~\cite{Garcia-Garcia:2000lra} which is qualitatively similar to what we 
observe for $T>T_c$ in Fig.~\ref{fig:ILM}.

\begin{figure}[ht]
\begin{center}
\includegraphics[scale=0.55]{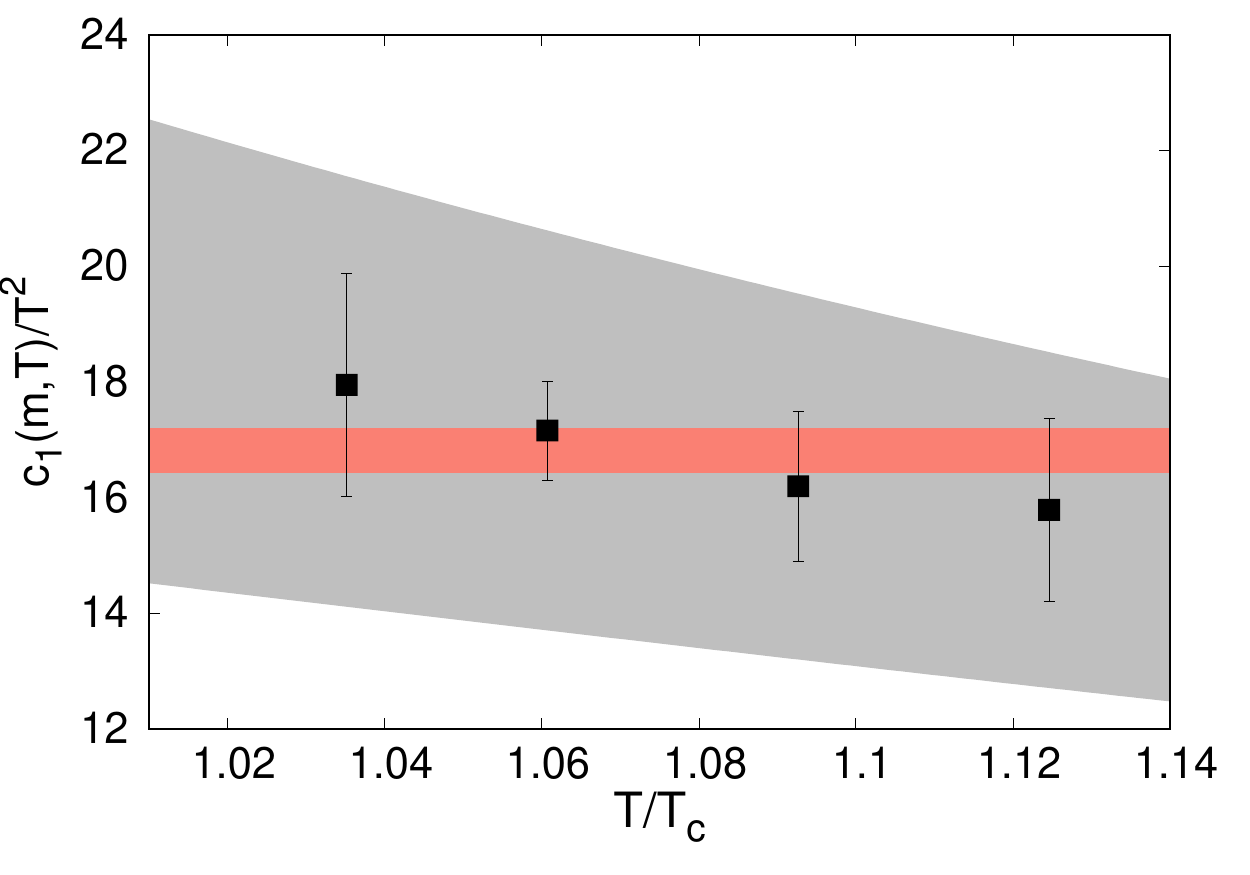}
\caption{Continuum estimates for $c_1(m,T)/T^2$ for $T> T_c$ 
obtained after fitting the points with an $m$-independent constant (orange band) and a sum of 
quadratic ($m^2/T^2$) and quartic ($m^4/T^4$) dependent fit function (gray band). }
\label{fig:c1fit}
\end{center}
\end{figure}

Now focusing on the bulk modes, it was shown using chiral Ward identities that in the 
symmetry-restored phase, the sufficient condition for $U_A(1)$ restoration 
evident from the degeneracy of up to six-point correlation functions in the scalar-pseudo-scalar 
sector are $c_1=\mathcal{O}(m^2)+...$ and $c_3=c_{30}+\mathcal{O}(m^2)+...$. 
The perturbative $\lambda^3$ growth in Eq.~\ref{eqn:evfit} can have a mass-independent 
coefficient which however does not lead to $U_A(1)$ breaking. We verify whether indeed 
it is true even non-perturbatively by performing a fit to the bulk part i.e. all 
eigenvalues $\lambda > \lambda_0$ with 
$\frac{\rho(\lambda)}{T^3} = \frac{\lambda}{T}.\frac{c_1(T,m)}{T^2}+\frac{\rho_0}{T^3}$. 
This ansatz neglects higher powers in $\lambda$ which is well justified since 
we are in the deep infrared of the eigenspectrum, represented by $\mathcal{O}(100)$ 
eigenvalues out of a total million available on such lattice sizes.  The results 
of the fit are discussed in Table~\ref{tab:fitparam}.  The extracted slope $c_1$ for each 
temperature $T>T_c$, at three different values of $N_\tau$ then allows us to perform 
a continuum ($\sim 1/N_\tau^2$) extrapolation of this coefficient. We next study the $m$ dependence 
of this continuum extrapolated coefficient $c_1(m,T)$. The results of the fits 
are shown in Fig.~\ref{fig:c1fit}. It is evident from the fit that it is more favorable that 
$c_1$ is proportional to $T^2$ ($\chi^2$/d.o.f=$0.6$) to leading order rather than 
$c_1$ is proportional to $m^2$ ($\chi^2$/d.o.f=$0.1$). From the fit we obtain the value 
of $c_1(m,T)/T^2=16.8(4)$.

The finite result for the slope in the continuum limit, i.e., the $m$-independent term in $c_1$ ensures 
that the $U_A(1)$ part of the chiral symmetry will remain \emph{minimally} broken in the chiral limit 
in the symmetry-restored phase as the maximum contribution to $U_A(1)$ breaking comes from the near-zero
eigenvalues, which we observe in the next section. Moreover the slope of the eigendensity for 
$T\lesssim 1.12~T_c$ is distinctly different from the perturbative $\lambda^3$ rise implying 
significant non-perturbative effects.

 
\begin{table}[h]
          \begin{tabular}{|c|c|c|c|c|c|}\hline
           $T$ [MeV]   &$N_\tau$ & $\lambda_0 /T$ &$\frac{c_1}{T^2}$ & $\rho_0/T^3$  \\ \hline\hline
             145& 12 & 0.1 & 9.0(5) & 7.30(7)\\\hline
             145& 16 & 0.05 & 9(1) & 6.67(9)\\\hline\hline
             162& 8 & 0.2 & 8.8(3) & 4.1(1)\\\hline
             162&12 & 0.15 & 13.2(2) & 2.69(5)  \\\hline
             162&16 & 0.1 & 17.5(5) & 1.93(7)  \\\hline\hline
             166 &8 & 0.2 & 8.9(1) & 3.31(5)  \\ \hline
             166 &12 & 0.15 & 13.3(3) & 1.92(6) \\ \hline
             166 &16 & 0.1 & 16.6(8) & 1.4(1)  \\\hline\hline
             171 &8  & 0.2 &9.3(1) & 2.38(5)  \\ \hline
             171 &12 & 0.15 & 12.9(1) & 1.19(3)  \\ \hline
             171 &16 & 0.1 & 17.0(5) & 0.45(8)  \\\hline\hline
             176 &8  & 0.2 &9.5(1) & 1.67(4)  \\ \hline
             176 &12 & 0.15 & 13.0(2) & 0.36(6)  \\ \hline
             \end{tabular} 
    \caption{Lattice sizes ($N_\sigma^3\times N_\tau$), 
    temperatures ($T$), the estimated values of $c_1/T^2$ and $\rho_0/T^3$ after the fit to the bulk modes which are defined beyond the lower cutoff at $\lambda_0/T$. }
    \label{tab:fitparam}
\end{table}

\subsection{The fate of $U_A(1)$ breaking in the continuum limit}

\begin{figure}[h]
\begin{center}
\includegraphics[scale=0.6]{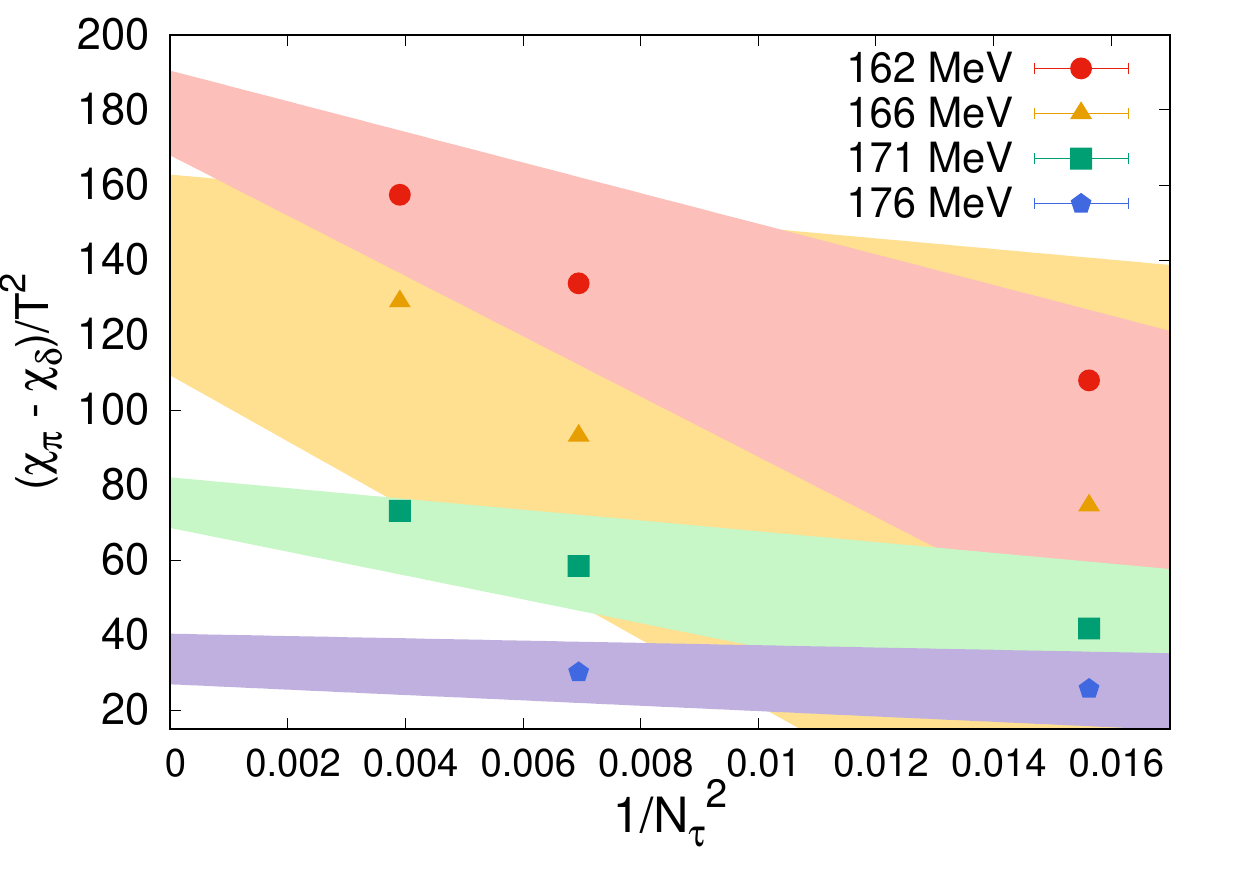}
\caption{The continuum estimates for $\chi_\pi-\chi_\delta$ normalized by the square of temperature 
for HISQ fermions from three different lattice spacings corresponding to $N_\tau=8, 12, 16$ at 
$T=162,166, 171$ MeV respectively and from $N_\tau=12, 16$ data at $T=176$ MeV. }
\label{fig:chipd}
\end{center}
\end{figure}

Since the flavor singlet part of the chiral symmetry is anomalous it has no corresponding 
order parameter. Hence to measure whether this singlet part of the chiral symmetry is 
simultaneously (and effectively) restored along with the non-singlet part, it has been 
suggested~\cite{Shuryak:1993ee} to look at the degeneracies of the integrated correlators
of mesons i.e., $\chi_\pi-\chi_\delta$. In the continuum, the integrated meson correlators 
are related to each other through the following relations, 
$\chi_\delta = \chi_\sigma - 4\chi_{\text{disc}}$ 
and $\chi_\pi = \chi_\eta + 4\chi_{5\text{disc}}$. These integrated meson
correlators are defined as 
$\chi_\pi = \int  d^4x ~\langle \pi^i(x) \pi^i(0)\rangle $, $\chi_\sigma = \int d^4x~ \langle 
\sigma(x) \sigma(0) \rangle $, $\chi_\delta = \int d^4x ~\langle \delta^i(x) \delta^i(0)\rangle 
$ and $\chi_\eta = \int d^4x~ \langle \eta(x) \eta(0)\rangle~\text{where}~i=1,2,3$. 
We measure $(\chi_\pi-\chi_\delta)/T^2$ at the four different temperatures 
above $T_c$, and perform a $\sim 1/N_\tau^2$ continuum extrapolation at each temperature, 
the results of which are shown in Fig.~\ref{fig:chipd}. For the highest temperature, we have 
only two data points available corresponding to $N_\tau=8, 12$ for performing the continuum extrapolation. We hence 
assigned $40 \%$ and $20 \%$ error to the values for the slope and the intercept respectively, similar 
to that obtained from a fit to the $T=171$ MeV data. This observable receives 99\% contribution
from the near-zero eigenvalues for $N_\tau=16$.  Performing continuum estimates with 
only two data points corresponding to finer lattice sizes $N_\tau = 16, 12$ at each 
temperature, gives a higher intercept than the corresponding extrapolation considering 
all three $N_\tau$ values. Hence the finiteness  of this observable is quite robust and 
we conclude that $U_A(1)$ does not get \emph{effectively} restored at $T_c$. 

\begin{figure}[h]
\begin{center}
\includegraphics[scale=0.6]{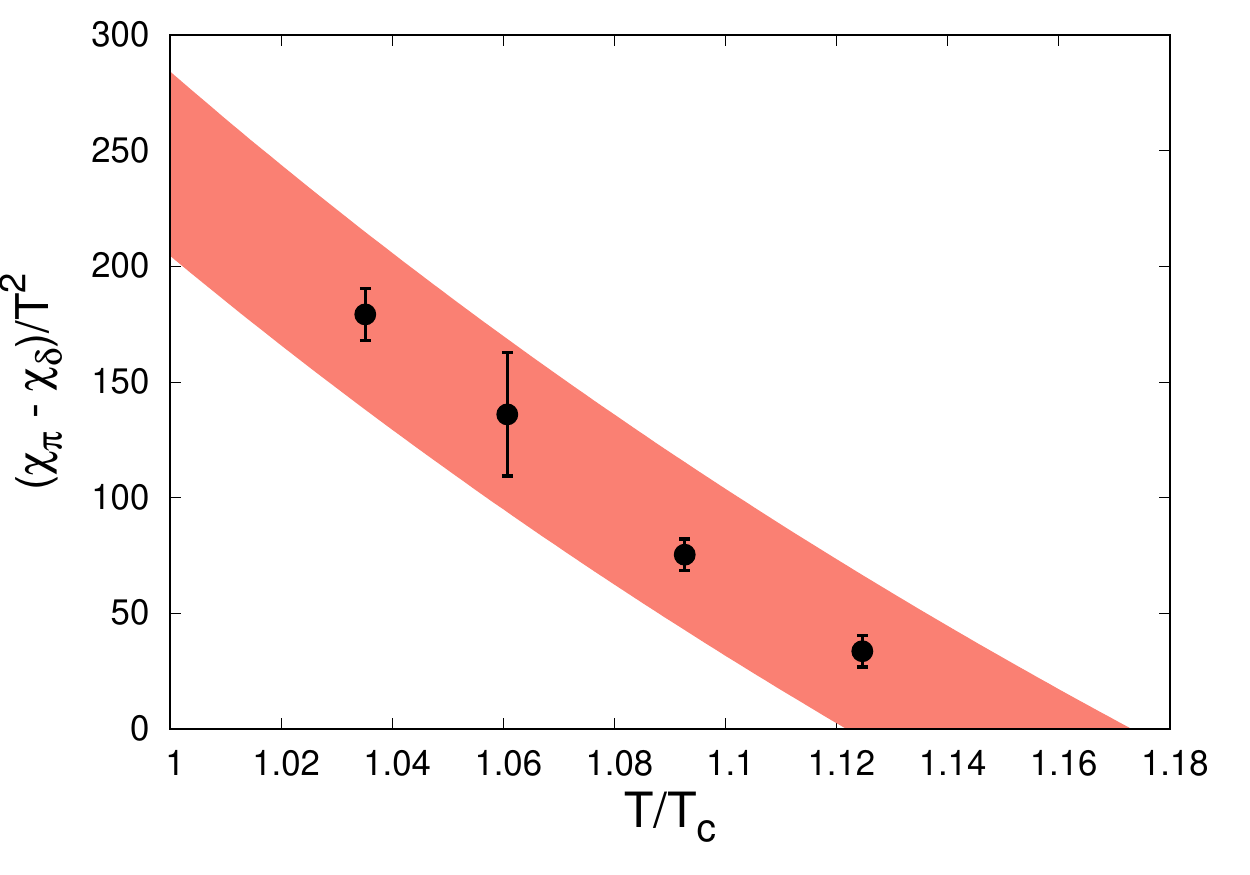}
\caption{The continuum estimates for $(\chi_\pi-\chi_\delta)/T^2$ for temperatures 
greater than $T_c$ shown as points fitted to a functional form  $a+b/T^2$ shown as 
a band.}
\label{fig:chipcdvsT}
\end{center}
\end{figure}

Motivated from the fact that the major contribution to $(\chi_\pi-\chi_\delta)/T^2$
comes from the near-zero modes, we expect a $1/T^2$ dependence to this quantity if the 
corresponding eigenvalue density can be characterized as a well-defined peak. Furthermore 
from the chiral perturbation theory at finite temperature one expects a similar 
$1/T^2$ dependence~\cite{Bonati:2015vqz} near $T\gtrsim T_c$. We thus fit the continuum extrapolated 
values of $(\chi_\pi-\chi_\delta)/T^2$ at each temperature $T>T_c$, i.e., the intercept 
of the fits shown in Fig.~\ref{fig:chipd} to the ansatz $a+b/T^2$. After performing the 
fit, shown in Fig.~\ref{fig:chipcdvsT} we extract the temperature $T/T_c=1.147(25)$ beyond 
which this $U_A(1)$-breaking observable drops to zero.

 Earlier analytic works based on the properties of integrated two-point meson 
 correlators argued that the $U_A(1)$ breaking comes from the eigenmodes of the
 Dirac operator at $\lambda=0$~\cite{Lee:1996zy} or close to zero~\cite{Cohen:1996ng}. 
 Whereas exact zero modes do not contribute in the thermodynamic limit, the density 
of near-zero modes at $T>T_c$ was expected to be zero~\cite{Cohen:1996ng} 
in the chiral symmetry-restored phase of QCD, whereas we observe a robust 
presence of the near-zero modes in the continuum limit which dominantly 
contributes to $U_A(1)$ breaking for $T\lesssim 1.15~T_c$.

We next compare our result with the earlier observation of $U_A(1)$ (effective) 
restoration temperature of $\sim 200$ MeV obtained from the continuum extrapolated results 
for the integrated screening correlators $(\chi_\pi-\chi_\delta)/T^2$ in 
$2+1$ flavor QCD using HISQ discretization~\cite{Bazavov:2019www} with heavier 
than physical light quark mass, corresponding to $m_\pi=160$ MeV. The corresponding pseudo-critical 
temperature is also higher by $\sim 4$ MeV; hence the restoration temperature comes out to be 
around $1.2~T_c$ which agrees with us within the error bar. Furthermore a recent 
work~\cite{Ding:2020xlj} has reported breaking of $U_A(1)$ due to a $m^2\delta(\lambda)$
feature in the eigenvalue spectrum at about $1.5~T_c$ which survives even in the chiral 
limit. We note here that $U_A(1)$ breaking due to this specific feature of the Dirac 
spectrum is expected to survive even at asymptotically high temperatures where the QCD 
vacuum can be explained in terms of a dilute gas of instantons~\cite{Dick:2015twa,Petreczky:2016vrs}. 
We however discuss here how a non-trivial breaking of the $U_A(1)$ part of chiral symmetry 
can arise due to features in the infrared part of the eigenvalue spectrum unlike the 
dilute instanton gas regime and show that beyond $1.15~T_c$ such a contribution gets 
effectively restored which then may transition into the temperature regime studied 
in Ref.~\cite{Ding:2020xlj}. We next verify that the chiral Ward 
identities are satisfied by the HISQ action.

\subsection{Verifying the chiral Ward identities}

\begin{figure}[h]
\begin{center}
\includegraphics[scale=0.6]{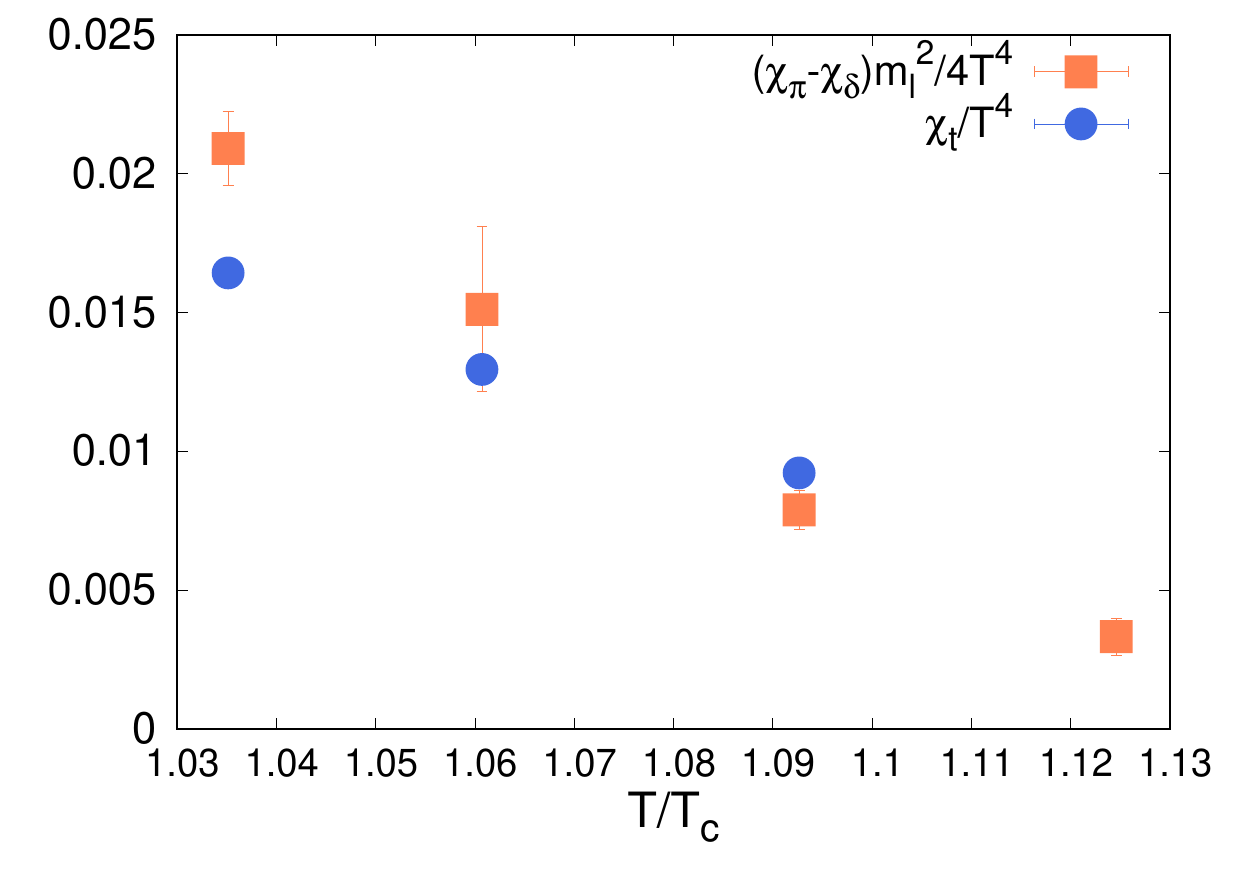}
\caption{A comparison of the integrated renormalized correlator 
$( \chi_\pi - \chi_\delta )m^2/4T^4$ with the topological susceptibility 
(measured independently using gradient flow in Ref.~\cite{Mazur:2021zgi}) 
for temperatures $>T_c$. }
\label{fig:chitopcomp}
\end{center}
\end{figure}

In the chiral symmetry restored phase, $\chi_\sigma = \chi_\pi$ and $\chi_\delta = \chi_\eta$ 
hence one obtaines $\chi_\pi - \chi_\delta = 4\chi_{5,\text{disc}}$. Using chiral Ward identities 
it is known that $\chi_{5,\text{disc}} = \chi_t/m^2$ where $\chi_t$ is the topological susceptibility 
of QCD. This allows one to relate the $U_A(1)$ breaking parameter to the topological susceptibility 
through the relation $1/4( \chi_\pi - \chi_\delta )m^2/T^4 = \chi_t /T^4$.  A comparison 
of these two observables is shown in Fig.~\ref{fig:chitopcomp}. From the figure it is evident 
that for $T>1.05~T_c$, when  chiral symmetry is effectively restored, the two quantities agree 
with each other within errors. This is particularly interesting since for staggered quarks, even 
though the chiral and taste symmetries are intermixed at finite lattice spacing, the symmetries of 
QCD and related chiral Ward identities are recovered in the continuum limit.

\subsection{Distribution of the lowest eigenvalue at finite temperature}

\begin{figure}[ht]
\begin{center}
\includegraphics[scale=0.6]{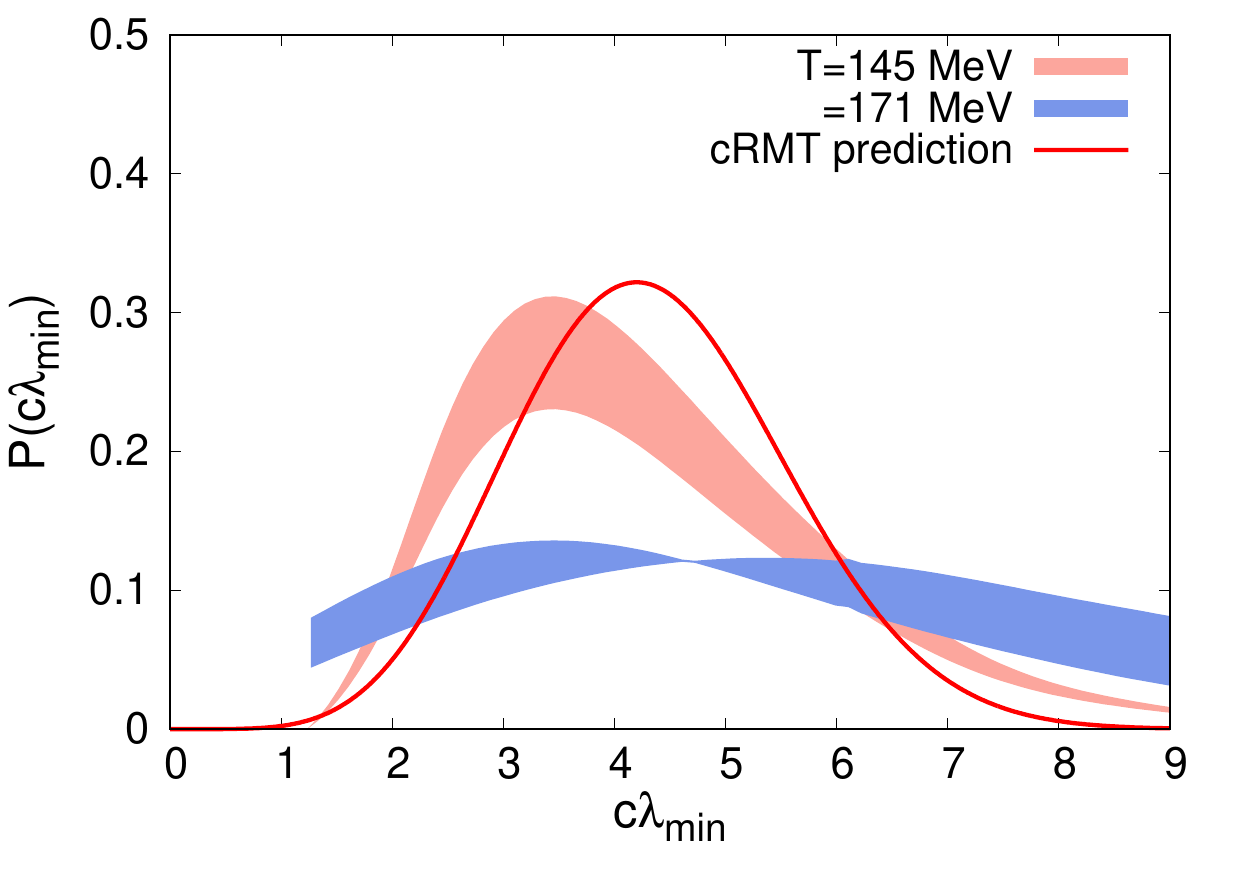}
\caption{The continuum extrapolated probability distribution of scaled lowest eigenvalue for $T=145,171$ MeV 
shown as orange and blue bands respectively and these are compared with the cRMT prediction for $N_f=2$. }
\label{fig:smallT}
\end{center}
\end{figure}

The probability distribution of the lowest eigenvalue of the QCD Dirac operator 
$\lambda_{\text{min}}$  has inherent information about the microscopic degrees of 
freedom. For a chiral random matrix ensemble for $N_f=2$ (at zero temperature) the 
lowest eigenvalue is distributed according to~\cite{Akemann:2016keq}
\begin{equation}
\label{eqn:rmtzmode}
P(x) = \frac{x}{2}e^{-\frac{x^2}{4}}\left[ J^2_2(x)-J_3(x)J_1(x) \right]~,~ x=c\lambda_{\text{min}} ~.~
\end{equation}
At the lowest temperature $T=145$ MeV, we calculate the probability distribution of the 
scaled lowest eigenvalue $ c \lambda_{\text{min}}$ at different lattice spacings and perform a 
continuum estimate of the distributions, 
for which we have extracted the lowest eigenvalue from each configuration for $N_\tau = 12,16 $
for $T=145$ MeV and $N_\tau = 8,12,16$ for $T=171$ MeV and later
rescaled to the dimensionless quantity $c\lambda_{\text{min}}$, where the value of 
$\langle \bar{\psi} \psi \rangle$ at finite temperature is obtained from 
Ref.~\cite{Steinbrecher:2018jbv}. Keeping the bin size constant 
we obtained the probability distribution of $c\lambda_{\text{min}}$ for each $N_\tau $ 
and then performed a spline interpolation by taking appropriate weights proportional to 
the errors for each data point in order to have a smoother interpolating curve. Next we 
performed a continuum extrapolation at each value of $c\lambda_{\text{min}}$ of the 
interpolating function with the ansatz $c + d/N_\tau^2$. We assigned a $15\%$ error 
for $T=145$ MeV, as we only had two points while performing the continuum extrapolation.
In order to compare the probability distribution of the lowest eigenvalue for both the temperatures 
with Eq.~4 we have to match their first moments with the cRMT distribution. We have 
calculated the first moment for Eq.~4 and found the result to be $4.344$. Next  
we have scaled the $c\lambda_{\text{min}}$ and the probability distribution for the 
lowest eigenvalue obtained at both these temperatures from our calculations in $2+1$ 
flavor QCD such that the first moment is exactly $4.344$ and the area under the curve is unity.
The probability distribution of the scaled lowest eigenvalue is shown in Fig.~\ref{fig:smallT}. 
The continuum extrapolation of the probability distribution at $T = 145 $ MeV shown as the orange 
band is close to the probability distribution of a $N_f = 2$ chiral Gaussian unitary random matrix 
ensemble. In contrast, we also plot the probability distribution of the scaled lowest eigenvalue 
at $T=171$ MeV whose continuum extrapolation is shown as a blue band in Fig.~\ref{fig:smallT}. 
It is evident that the lowest eigenvalue which is a part of the near-zero peak follows a very 
different statistics rather than known from a chiral RMT.

\subsection{The level spacing distribution for bulk modes}

\begin{figure*}[t]
\begin{center}
\includegraphics[scale=0.34]{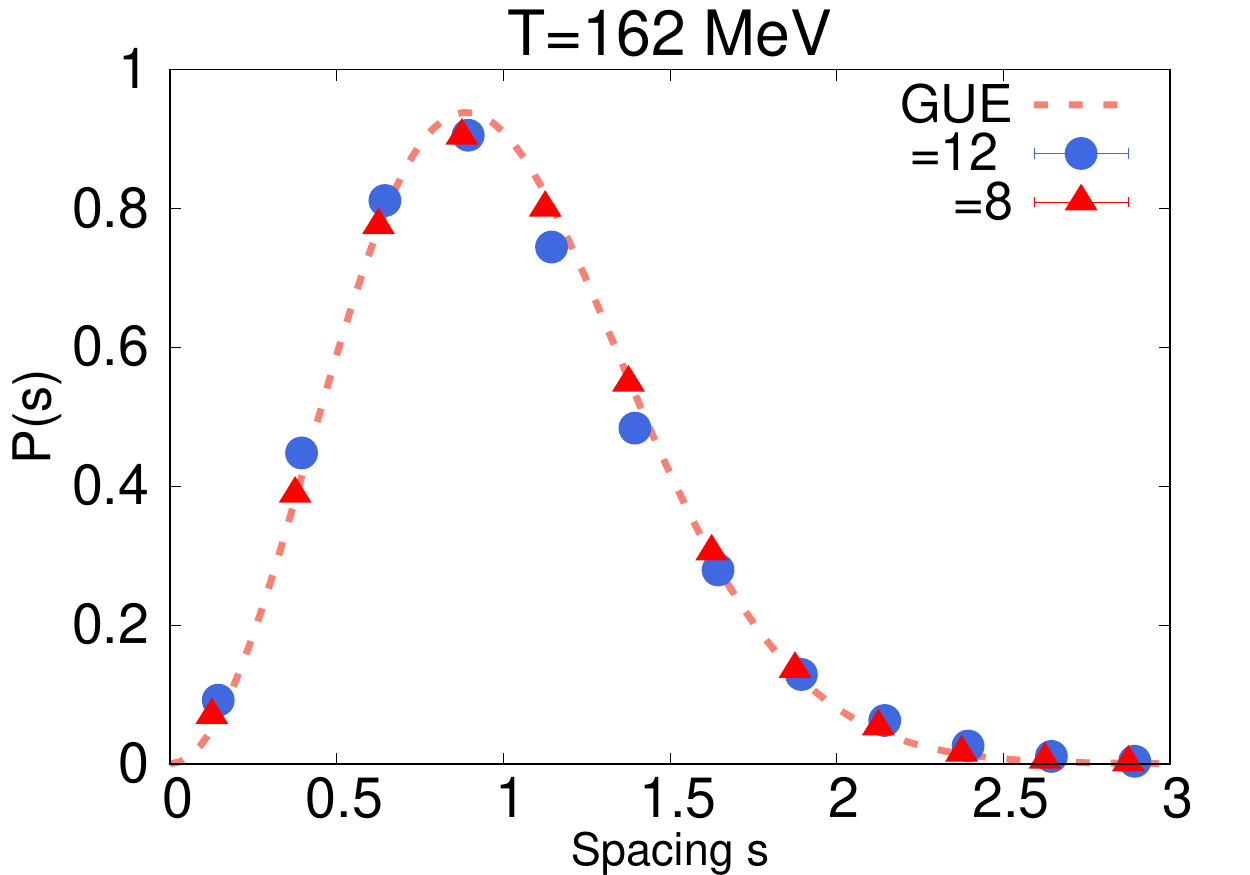}
\includegraphics[scale=0.34]{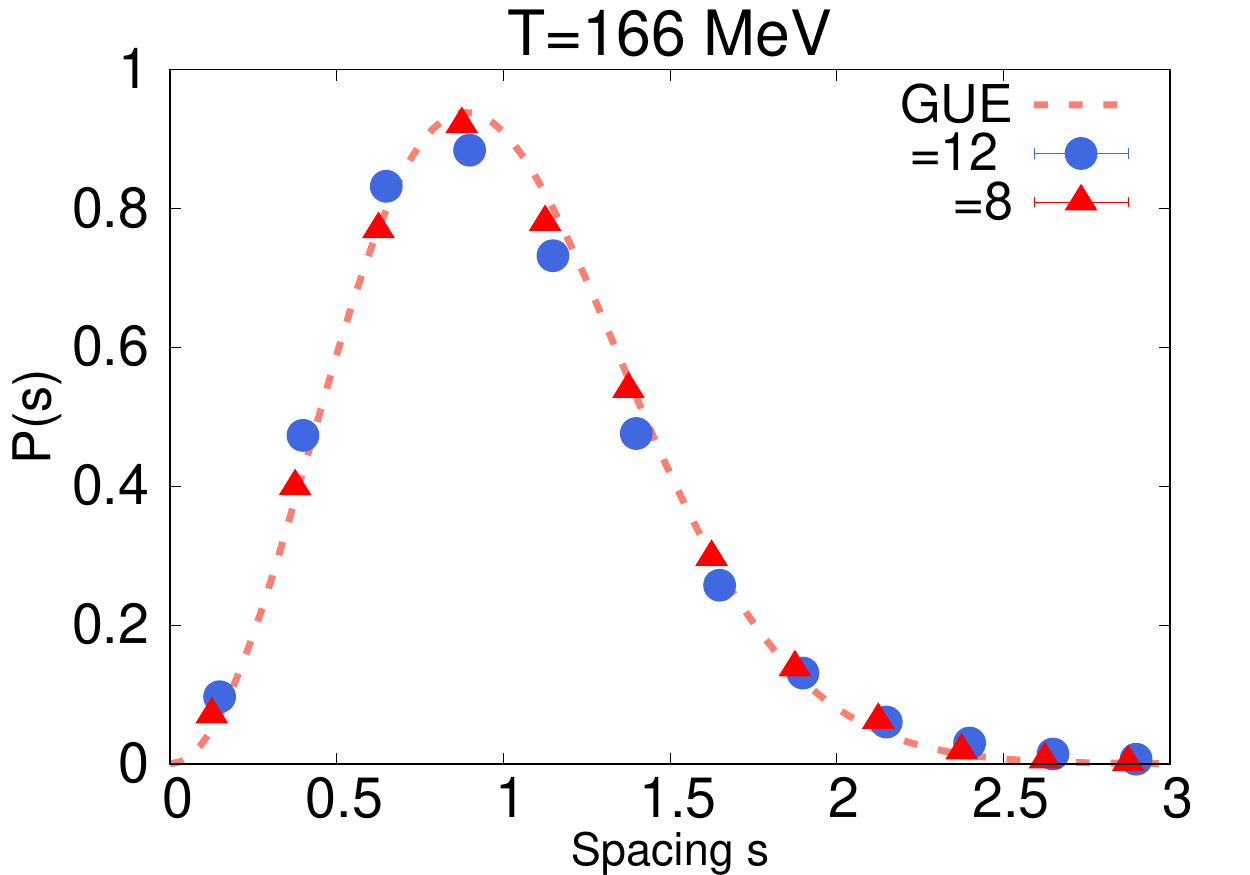}
\includegraphics[scale=0.34]{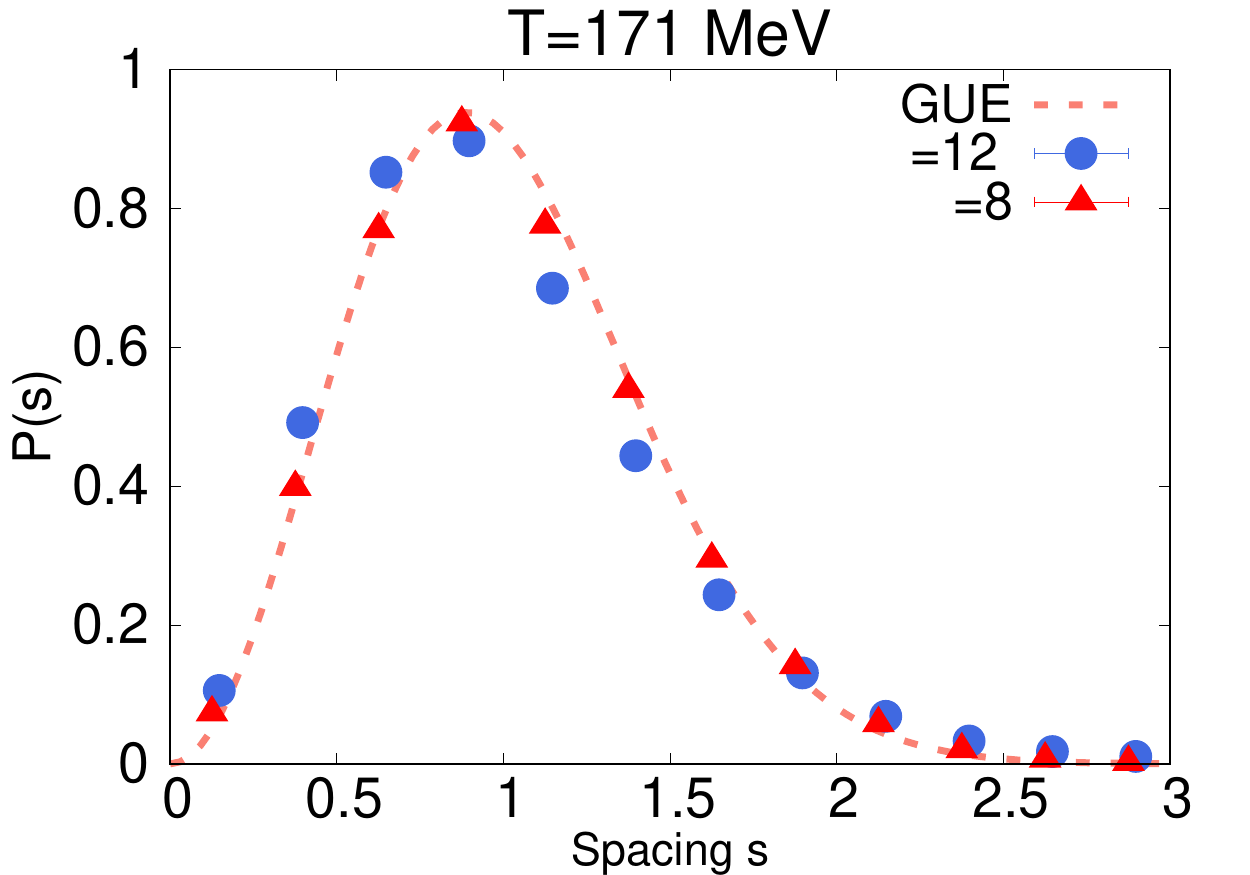}
\includegraphics[scale=0.34]{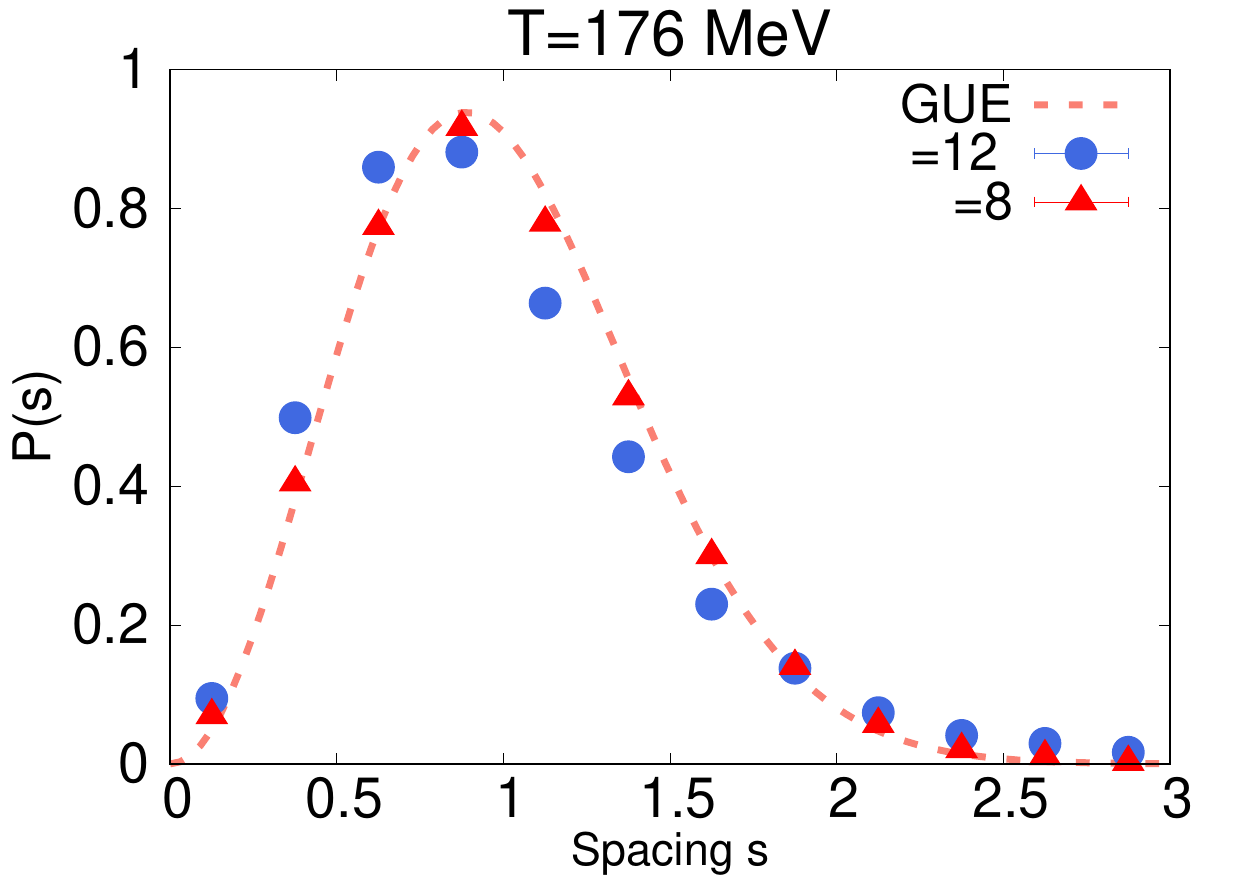}
\caption{Unfolded level spacing distributions of bulk eigenvalue modes for different temperatures shown as a 
function of different lattice spacings or equivalently, $N_\tau$. The dotted curves in each plot correspond 
to the Wigner surmise for Gaussian unitary random matrix ensembles.}
\label{fig:unfold}
\end{center}
\end{figure*}

\begin{figure}[h]
\begin{center}
\includegraphics[scale=0.6]{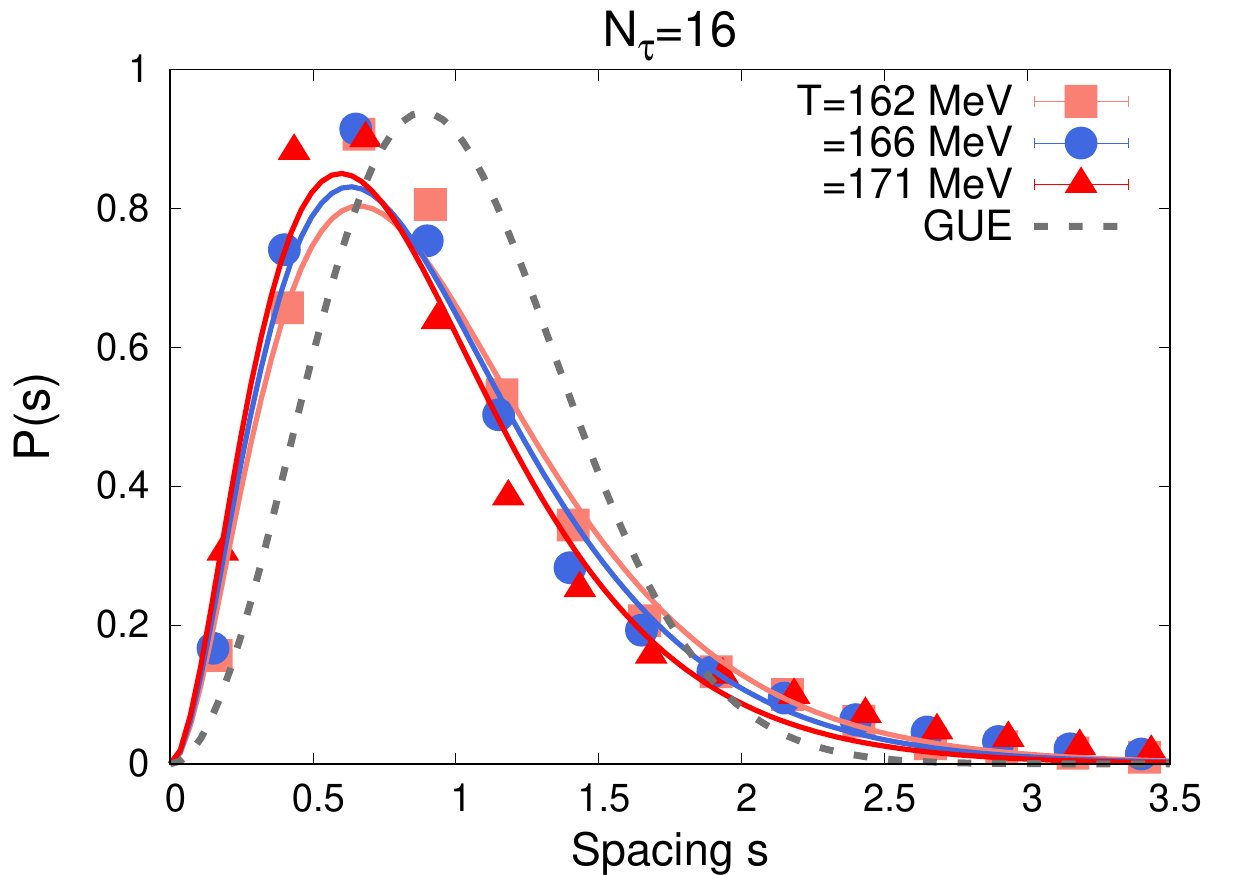}
\caption{A fit to the eigenvalue level spacing distribution using a mixed ansatz for 
$N_\tau = 16$ at $T = 162, 166, 171$ MeV and compared with the prediction from a GUE of random matrices.}
\label{fig:unfold-semi}
\end{center}
\end{figure}

In order to understand the properties of bulk modes we look at their 
level spacing distribution. To study the universal properties of the 
eigenvalue level spacing fluctuations one has to remove the system 
dependent mean. This is done by a method called unfolding. 
Let $\lambda$ represent eigenvalues in the ascending sequence for any particular gauge configuration. 
The average density of the eigenvalues in the sequence i.e. the 
reciprocal of the average spacing as a function of $\lambda$ is represented as 
$\bar{\rho} (\lambda)$. The eigenvalue sequence can then be unfolded using the average 
level-staircase function,
$\bar{\eta}(\lambda) = \int_{\lambda_0}^{\lambda}~ d\lambda ' \bar{\rho}(\lambda ')$ 
which tells us how many eigenvalues in this sequence are less than $\lambda$ on an average. 
Here $\lambda_0$ labels the eigenvalue beyond which all the higher 
eigenvalues are bulk modes and below which are the near-zero modes.
The unfolded sequence is labeled by  $\lambda^{uf}_i = \bar{\eta}(\lambda_i)$, where the 
index $i$ labels the original eigenvalue whose unfolding is performed. When 
appropriately normalized, the average spacing between the unfolded eigenvalues equals unity.
The nearest neighbor spacing distribution is constructed by calculating the differences 
between consecutive unfolded eigenvalues $\lambda^{uf}_{i+1}-\lambda^{uf}_i$ and organizing them
into histogram bins. This gives us a picture of how the 
eigenvalue spacings fluctuate about the average which we have shown in Fig.~\ref{fig:unfold} 
for four  different temperatures $T=162, 166, 171, 176$ MeV and at each temperature, for the two 
different lattice sizes $N_\tau = 8,12$. We then compare the nearest neighbor spacing distributions
to the Wigner surmise for the Gaussian unitary ensemble (GUE) given by $P(s)= 32/\pi^2 s^2 \rm {e}^{-4 s^2/\pi}$
shown as dotted lines in Fig.~\ref{fig:unfold}. It is evident that the level repulsion between 
the bulk modes for small $s$ is quadratic similar to that of random matrices belonging to the GUE.
We see however that as the temperature increases, the agreement to GUE with the $N_\tau = 12$ data 
for $s>1$ is not so good, whereas the low $s<1$ part agrees very well. This occurs due to the 
contamination of the bulk modes, which are more closely spaced than the near-zero modes which 
start to build up forming a peaklike structure in the infrared part of the eigenvalue spectrum.
For the $N_\tau=16$ lattices which has a clear well-defined peak of near-zero modes at $T\geq 166$ 
MeV, the contamination with the bulk modes is expected to be even more severe. As expected the 
comparison of the tail of the spacing distribution of the $N_\tau=16$ data to the GUE prediction 
produces not-so-good agreement. In order to account for the long tail of the spacing distribution 
we fit it to a distribution $P(s)\sim s^2\exp{(-\alpha s)}$ which shows strong quadratic level repulsion 
at small values of $s$ but falls off slowly at large values of $s$ parametrized by a fit parameter 
$\alpha$. After performing the fit of the level separation with this ansatz, we obtain the value of
$\alpha = 3.02(7), 3.17(9), 3.3(1)$ for temperatures $T = 162, 166, 171$ MeV respectively.
The lattice data now do agree to this new fit ansatz reasonably well for  $N_\tau=16$ at
all temperatures above $T_c$, which is evident in Fig.~\ref{fig:unfold-semi}.

This is a generic feature of the eigenvalue spacing distribution of a strongly disordered 
system~~\cite{BASKO20061126} whose bulk eigenmodes in the center of the band follows a 
similar behavior as RMTs except for the tails of the distribution due to contamination 
with the localized states. We will explain this feature in more detail in the next 
section.

\subsection{Separating the near-zero from the bulk eigenvalues of the QCD Dirac spectrum}

\begin{figure}[hb]
\begin{center}
\includegraphics[scale=0.6]{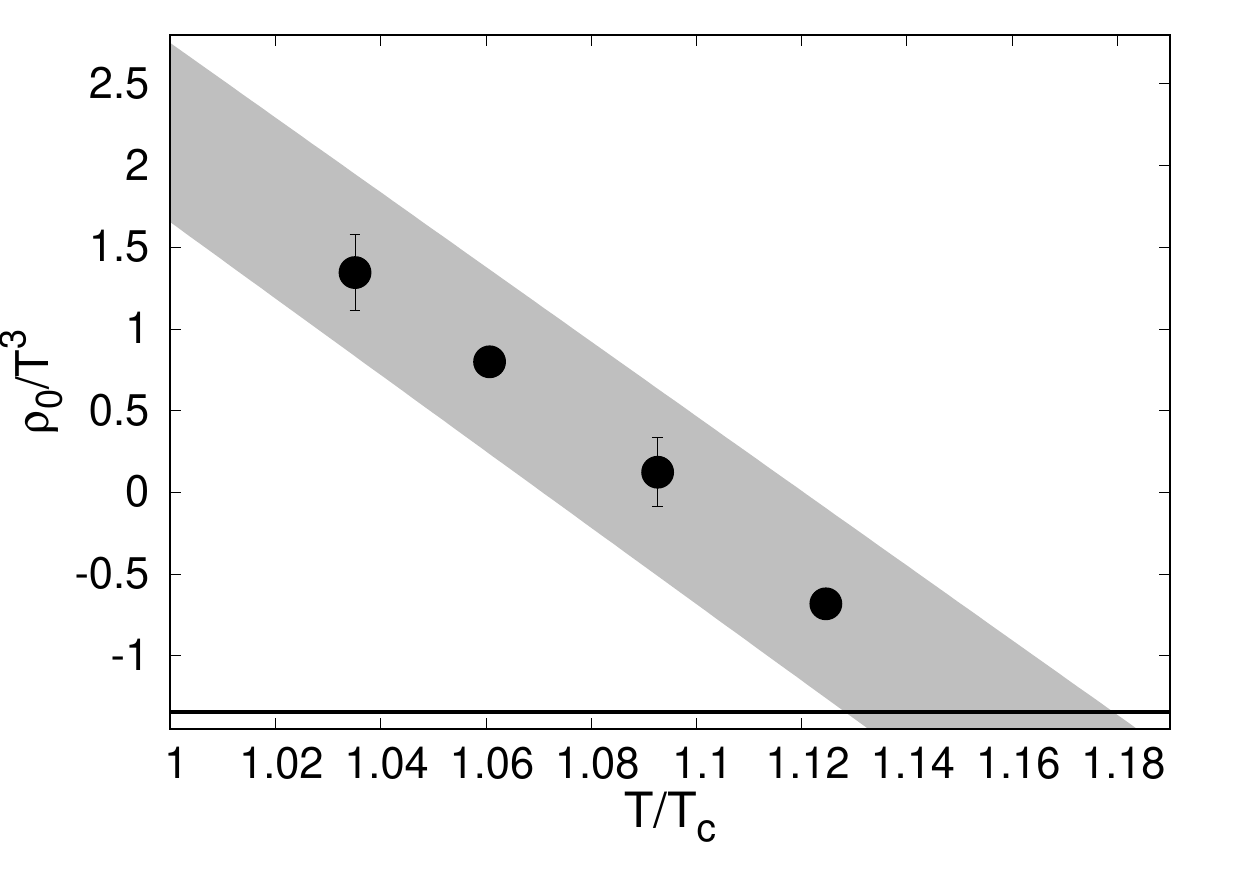}
\caption{Continuum extrapolation of the bulk intercept for eigenvalue densities at 
different temperatures above $T_c$. The horizontal line corresponds to $\rho_0/T^3 = -1.34$ 
for the bulk spectrum when it completely separates from near-zero modes.}
\label{fig:AndersonL}
\end{center}
\end{figure}

Having shown the distinct features of near-zero and bulk modes, the 
next question we ask is whether the near-zero modes which arise due to 
instanton interactions can distinctly disentangle out from the 
delocalized bulk modes. The QCD medium above $T_c$ consists of quarks 
interacting with each other as well residing in a disorder potential
very similar to an interacting electron system in a background random 
potential studied in detail in Ref.~\cite{BASKO20061126}. Such systems 
have a mixing between the localized and delocalized many-body states which 
is in contrast to the traditional Anderson model, consisting of non-interacting 
electrons in the presence of a random disorder. In the Anderson model, one-electron 
states with the same energy but with different localization properties cannot 
co-exist in three or more spatial dimensions, $d\geq 3$. There exist bands of localized and 
extended states, and a unique energy separating two such bands for $d\geq 3$ which 
is termed as a mobility edge. The QCD medium above $T_c$ however cannot be understood through 
a conventional Anderson model; it has far more interesting properties, like 
the possibility of the existence of a scale-invariant infrared phase above $T_c$ 
discussed in the recent literature~\cite{Alexandru:2021pap,Kehr:2023wrs}. In fact 
we do observe a mixing between the localized states with the bulk spectrum of 
the QCD Dirac operator in the level spacing distribution data as shown in 
Fig.~\ref{fig:unfold-semi}.


In order to estimate the temperature when the bulk modes separate out from the 
deep-infrared peak of eigenvalues, we first estimate the typical spread of the 
near-zero peak visible in the eigenvalue density plots corresponding to $N_\tau=16$ 
in Fig.~\ref{fig:hisqev}. Recall that we have already estimated the slope and 
the intercept of the bulk eigenvalue density, using which we subtract the bulk 
mode contribution from the total eigenvalue spectrum for the $N_\tau=16$ data at 
each temperature. The near-zero peak which we get after subtracting the bulk 
modes has a typical spread which we estimate to be $\lambda/T=0.08$ for all the 
temperatures above $T_c$. Next, using the fact that the bulk modes have a 
linear-in-$\lambda$ dependence with a slope $c_1/T^2=16.8(4)$ 
in the continuum and the near-zero and bulk modes will separate out at 
a particular temperature, leading to the density to drop to zero at 
$\lambda/T=0.08$, we estimate the value of the bulk intercept $\rho_0/T^3=-1.34$ 
(corresponding to $\lambda=0$) in the continuum limit. 
We then take the values of the intercept $\rho_0/T^3$ of bulk mode density for all $T>T_c$ 
from Table~\ref{tab:fitparam} and perform a continuum extrapolation with an ansatz 
$\rho_0/T^3+ d/N_{\tau}^2$. The continuum values of the quantity $\rho_0/T^3$ for $T>T_c$, 
so-obtained after the fit are shown in Fig.~\ref{fig:AndersonL}.  At the highest temperature 
$T=176$ MeV, a $10\%$ error is assigned to the data point since we could perform a 
continuum estimate, with data available only for two $N_\tau$ values. Next, fitting the 
continuum extrapolated data for $\rho_0/T^3$ as a function of temperature with a fit 
ansatz $\rho_0/T^3=d_1(T/T_c)+d_2$ we obtain the fit parameters to be $d_1=-23.1(3)$ 
and $d_2=25.3(3)$ respectively. After obtaining this parametric dependence of the 
continuum estimates of the intercept as a function of temperature, we can now 
extract the temperature where the value of the intercept $\rho_0/T^3 = -1.34$, i.e., 
when the near-zero modes distinctly emerge out from the bulk spectrum. The extracted 
temperature comes out to be $T=1.15(3)~T_c$. This is within the temperature range 
when the $U_A(1)$ part of the chiral symmetry is \emph{effectively} restored.

\section{Why is $U_A(1)$ effectively restored at temperatures above $T_c$?}

In order to interpret these results, one could visualize quarks as a many-body state
moving in the background of an interacting ensemble of instantons, where the strength 
of the interactions changes as a function of temperature. At the microscopic level 
it is conjectured that the instantons remain strongly correlated below 
$T_c$, subsequently transitioning to a liquid-like phase with a finite but weaker correlation 
length~\cite{Schafer:1996wv} just above $T_c$, and eventually to a gas-like phase around $2~T_c$~
\cite{Petreczky:2016vrs,Ding:2020xlj}. Below $T_c$ the intercept of the infrared eigenvalue 
density quantifies the chiral condensate which corresponds to the breaking of 
the non-singlet part of the chiral symmetry. Owing to very strong correlations the 
microscopic details of the interactions are lost and the eigenvalues repel strongly 
similar to random matrices of a GUE. As the temperature is increased, at $\sim 171$ MeV, 
the near-zero eigenvalues start to become prominent. These eventually separate from the bulk 
at $\sim 1.15~T_c$. Earlier studies have observed screening of inter-instanton interactions 
and build-up of local pockets of Polyakov loop fluctuations~\cite{Bruckmann:2011cc,Holicki:2018sms,Baranka:2022dib} 
above such temperatures. This is also the region where the constituent dyons of the 
closely-spaced instantons interact semi-classically and thus start to become 
detectable~\cite{Bornyakov:2015xao,Larsen:2018crg,Larsen:2019sdi,Larsen:2021ppf}. 

Incidentally this suppression of long range instanton interactions also weakens the 
strength of $U_A(1)$ breaking, allowing for its \emph{effective} restoration at 
$T\gtrsim 1.15~T_c$. Lattice studies~\cite{Amato:2013naa,Aarts:2014nba} have reported 
a jump in the electrical conductivity around this temperature. 
Similar phenomena have also been reported in many-electron systems~\cite{BASKO20061126} in a 
disordered potential where the interplay between disorder and interactions causes a separation between 
the localized and delocalized states leading to a jump in the electrical conductivity from near-zero 
to a finite value.

\section{Conclusions}
In this work we have addressed a long-standing question of whether the flavor singlet $U_A(1)$ 
subgroup of the chiral symmetry gets effectively restored simultaneously with the non-singlet part 
for QCD with two light quark flavors at $T_c$. The effective restoration of the anomalous $U_A(1)$ 
symmetry is a non-perturbative phenomenon driven by the deep infrared part of the QCD Dirac 
eigenvalue spectrum. By carefully performing the continuum extrapolation of the staggered 
Dirac spectrum on the lattice and studying in detail its properties, we explicitly demonstrate 
that  $U_A(1)$ remains effectively broken in the chirally symmetric phase ($T> T_c$) for
$T\lesssim 1.15~T_c$. We also provide arguments for why this conclusion should remain unchanged 
even in the chiral limit. 


With the increase in temperature the strength of interactions
between the instantons starts to weaken due to which the deep
infrared peak of the spectrum is separated out from the bulk modes, 
which happens at around $T\sim 1.15~T_c$. The tunneling probability 
due to instantons also decreases with increasing temperature which 
results in lowering of the height of the near-zero peak of eigenvalue density. 
We show for the first time that both these phenomena are possibly the reason 
behind the $U_A(1)$ restoration, which also surprisingly happens to be around 
the same temperature. Observations of such rich interplay of phenomena in 
QCD matter above $T_c$ should be quite robust, since these are made after performing 
a continuum extrapolation. It will be interesting to observe further finer details of 
chiral transition in the massless limit with QCD Dirac operators which have exact chiral 
symmetry on the lattice.

 All data from our calculations, presented in the figures of this paper, can be found in
Ref.~\cite{DOI}.

\textbf{Acknowledgements}
The authors acknowledge support by the Deutsche For\-schungs\-ge\-mein\-schaft 
(DFG, German Research Foundation) through the CRC-TR 211 'Strong-interaction matter 
under extreme conditions'– Project no. 315477589 – TRR 211. S.S. acknowledges support 
by the Department of Science and Technology, Government 
of India through a Ramanujan Fellowship. The numerical computations in this 
work were performed on the GPU cluster at Bielefeld University.
We thank the Bielefeld HPC.NRW team for their support.
We thank the HotQCD Collaboration, specially Christian Schmidt for
sharing the gauge configurations and software with us. We acknowledge 
the contribution of Hiroshi Ohno who was involved during the early stages of the 
project. S.S. is grateful to Frithjof Karsch for helpful discussions and his kind 
hospitality when this work was finalized. We also thank Gernot Akemann for his 
very helpful guidance and Matteo Giordano for his suggestions on the first draft. 
A part of this work is based on the MILC Collaboration’s public lattice gauge 
theory code~\cite{MILC}.

\bibliography{references_paper.bib}

\end{document}